\documentclass[preprint,amsmath,amssymb,aps,eqsecnum,prd,floatfix]{revtex4}
\usepackage{graphicx}
\usepackage{multirow}
\graphicspath{ {./} {fig/} }

\def \H{{\cal H}}

\def \D{\Delta}

\begin{document}


\title{Improving the efficiency of the detection of gravitational wave
  signals from inspiraling compact binaries: Chebyshev interpolation} 
\author{S.~Mitra}
\email{sanjit@iucaa.ernet.in}
\author{S.~V.~Dhurandhar}
\email{sanjeev@iucaa.ernet.in}
\affiliation{Inter-University Centre for Astronomy and Astrophysics,
  Ganeshkhind, Pune - 411 007, India} 
\author{L.~S.~Finn}
\affiliation{Center for Gravitational Wave Physics, 
The Pennsylvania State University, University Park, PA 16802, USA}
\altaffiliation{Also Institute for Gravitational Physics and Geometry,
  Department of Physics, and Department of Astronomy and Astrophysics} 
\email{LSFinn@PSU.Edu}
\date{\today}

\begin{abstract}

  Inspiraling compact binaries are promising sources of gravitational
  waves for ground and space-based laser interferometric detectors.
  The time-dependent signature of these sources in the detectors is a
  well-characterized function of a relatively small number of
  parameters; thus, the favored analysis technique makes use of
  matched filtering and maximum likelihood methods.
  As the parameters that characterize the source model are varied so
  do the templates against which the detector data are compared in the
  matched filter. For small variations in the parameters, the output
  of the matched filter for the different templates are closely
  correlated.
  Current analysis methodology samples the matched filter output at
  parameter values chosen so that the correlation between successive
  samples is 97\%.
  Correspondingly, with the additional information available with each
  successive template evaluation is, in a real sense, only 3\% of that
  already provided by the nearby templates.
  The reason for such a dense coverage of parameter space is to
  minimize the chance that a real signal, near the detection
  threshold, will be missed by the parameter space sampling.
  Here we describe a straightforward and practical way of using
  interpolation to take advantage of the correlation between the
  matched filter output associated with nearby points in the parameter
  space to significantly reduce the number of matched filter
  evaluations without sacrificing the efficiency with which real
  signals are recognized.
  Because the computational cost of the analysis is driven almost
  exclusively the matched filter evaluations, a reduction in the
  number of templates evaluations translates directly into an increase in
  computational efficiency.  Because the computational cost of the
  analysis is large, the increased efficiency translates also into an
  increase in the size of the parameter space that can be analyzed
  and, thus, the science that can be accomplished with the data.
  As a demonstration we compare the present ``dense sampling''
  analysis methodology with our proposed ``interpolation''
  methodology, restricted to one dimension of the multi-dimensional
  analysis problem.  We find that the interpolated search reduces by
  25\% the number of filter evaluations required by the dense search
  with 97\% correlation to achieve the same efficiency of detection
  for an expected false alarm probability. Generalized to the two
  dimensional space used in the computationally-limited current
  analyses this suggests a factor of two increase in computational
  efficiency; generalized to the full seven dimensional parameter
  space that characterizes the signal associated with an eccentric
  binary system of spinning neutron stars or black holes it suggests an
  order of magnitude increase in computational efficiency.


\end{abstract}


\maketitle

\section{Introduction}
 
Inspiraling compact binaries of stellar mass neutron stars or black holes 
are among the most important gravitational wave sources
accessible to the current generation of ground-based interferometric 
gravitational wave detectors \cite{sanders03a,LSC04c,ando00a,VIRGO03A}. 
They are also very ``clean'' systems, in the sense that the gravitational 
wave signal arising from the inspiral depends only on general relativity
(i.e., the structure of the binary components is unimportant) and can 
be calculated to great accuracy by the well-understood techniques of 
post-Newtonian perturbation theory \cite{blanchet95b,damour01a,damour02a}. 
For these reasons, matched filtering and maximum likelihood techniques
are well-suited for the detection and characterization of the signal
from these systems \cite{finn92a,finn93a} and an implementation based
on these methods is currently used in the analysis of data from the
LIGO and GEO detectors \cite{LSC04d}.

The gravitational wave signature of inspiral binary systems depends on
a set of 15 parameters that characterize the system (i.e., component
masses, orbital energy and angular momentum at a given epoch,
component spins, orientation relative to detector line of sight). To
identify an incident signal using a matched filter requires the
application of a fair sampling of filter ``templates'', each defined by
a unique choice of the parameters associated with the physical system. 
Current implementations of matched filtering used in the analysis of 
gravitational wave detector data involve a very dense sampling of the 
two-dimensional parameter subspace corresponding to the binary 
component masses (intrinsic parameter space) 
and assuming zero eccentricity orbits and no body 
spins\footnote{The rationale for choosing a subspace is that the 
computational cost of a full parameter space search is high and that many 
systems are believed to be adequately represented by this subspace. Even 
for this two dimensional subspace the minimum computational cost for a 
matched filter search over component masses in the range $0.2\,M_{\odot} <
m_1 \leq m_2 < 30\,M_{\odot}$ in the LIGO detector band is several
hundred GFlops/s \cite{owen96a}. When significant body spin is allowed 
the computational cost grows by several orders of magnitude 
\cite{buonanno03a}.}: the templates are spaced so closely that the 
correlation between templates at neighboring points in the subspace is 
97\% \cite{sathyaprakash91a,dhurandhar94b}.  

We refer to this as the ``dense'' search strategy.  The rationale underlying 
the dense search strategy is to reduce the probability that a weak signal, 
characterized by parameters that fall
between those sampled, will be missed by the sampling. Here we
describe a straightforward and practical way of using interpolation to
take advantage of the correlation between the matched filter output
associated with nearby points in the parameter space to significantly
reduce the number of matched filter evaluations without sacrificing
the efficiency with which real signals are recognized.

We are not the first to observe the significance of the high
correlation between neighboring templates nor to consider the
opportunity for and advantages of interpolation as part of the
implementation of matched filtering for the analysis of binary
inspiral signals. The significance of the high correlation as an
indication that fewer templates should be able to recover signals with
the same efficiency, was first made in \cite{dhurandhar94b}.
Croce et al \cite{croce00a,croce00b} explored the use of Cardinal
interpolation with a truncated sinc function to estimate the value of
the matched filter output when the filter used corresponds to the
actual parameters that describe the signal. They found a sampling of
parameter space that would insure the interpolated estimate would be
no less than 97\% of the maximum over a two dimensional intrinsic parameter
space.  Their sampling and interpolation reduced by a factor of 4,
compared to the dense search, the number of templates required to
search over a two dimensional intrinsic parameter space.  Here we find that we
can achieve the same increase in efficiency per parameter space
dimension with a simpler template spacing and a simpler and quicker to
evaluate interpolation function.

Other suggestions have been made for reducing the number of matched
filter evaluations without sacrificing detection efficiency. One
promising proposal involves a hierarchical search strategy, wherein a
low-threshold trigger generated by the evaluation of the matched
filters associated with a much coarser sampling of parameter space
followed by (if necessary) a higher threshold evaluation matched
filters over a much finer sampling of parameter space
\cite{mohanty96a,mohanty98a,croce04a,croce04b,sengupta03a}. The
interpolation proposal we make here is complementary in the sense that
it can be implemented together with the hierarchical strategies that
have already been proposed to further improve the computational
efficiency of binary inspiral analysis. While  
the gain in efficiency of the interpolated search over the dense  
search is approximately constant in the desired false alarm  
probability, the balance between the coarseness of the grids in the  
hierarchical steps, the number of hierarchical steps, and the gain in  
computational efficiency associated with the interpolation is not  
obvious and requires further study.

The paper is organized as follows: In section~\ref{gen_method} we 
describe the motivation behind our choice of interpolating function
and the difference between our choice and the choice 
made in  \cite{croce00a,croce00b}.
In section~\ref{sec:Comparison} we describe in detail the 
dense and interpolated search strategies, the two-dimensional 
template space used in current gravitational wave data analyses 
for inspiraling binary neutron stars, the one-dimensional 
restriction that we use here to compare the effectiveness of 
the interpolating search strategy, and (finally) 
compare the performance of the
interpolated and dense search strategies by
evaluating the sensitivity of each at fixed computational cost
and the computational cost required by each to achieve the same 
sensitivity. 

\section{Interpolating in parameter space}
\label{gen_method}


The Wiener matched filter $W$, corresponding to an expected signal
characterized by $\boldsymbol{\tau}$, is a scalar-valued function of the
(vector-valued) instrument data $\mathbf{d}$, noise power spectral
density $S_n$:
\begin{equation}
W(\mathbf{d}|\boldsymbol{\tau}) = W(\boldsymbol{\tau}|S_n,\mathbf{d}).
\end{equation}
In our particular problem $W(\mathbf{d}|\boldsymbol{\tau})$ is a continuous
function of $\boldsymbol{\tau}$ and $\boldsymbol{\tau}$ corresponds to
the parameters that characterize our binary system model: e.g., binary
system component masses, orbital energy and angular momentum,
component spins, etc. Given a data set $\mathbf{d}$ we wish to find 
an interpolating function
$\widetilde{W}(\boldsymbol{\tau})$ and a set of points 
$\boldsymbol{\tau}_k$ in the space of possible signals such that
\begin{equation}
W_k = \widetilde{W}(\boldsymbol{\tau}_k) =
W(\mathbf{d}|S_n,\boldsymbol{\tau}_k). 
\end{equation}
There are, of course, an infinite number of continuous functions
$\widetilde{W}(\boldsymbol{\tau})$ that take on the values $W_k$ at
the $\boldsymbol{\tau}_k$: the question is, how do we choose among
them?

Focus attention first on the case where $\boldsymbol{\tau}$ is a
scalar $x$. One particular choice of interpolant $\widetilde{W}(\mathbf{d}|S_n,x)$,
which is especially important in the context of communication theory,
is based on the Whittaker Cardinal function $\mathrm{sinc}$:
\begin{equation}
C(x) = 
\sum_{k=-\infty}^\infty
W_k \mathrm{sinc}\frac{x-x_k}{\Delta},
\end{equation}
where
\begin{eqnarray}
\mathrm{sinc}(x) &=& \frac{\sin\pi x}{\pi x},\\
x_k &=& x_0 + k\Delta.
\end{eqnarray}
Shannon \cite{shannon49a} showed that the Cardinal interpolation
$C(x)$ of ${W}(\boldsymbol{d}|S_n,x)$ is the unique interpolant $\widetilde{W}$ that
\emph{(i)} takes on the values $W_k$ at the $x_k$, \emph{(ii)} has no
singularities, and \emph{(iii)} and whose spectrum is limited to a
bandwidth $(2\Delta)^{-1}$. Correspondingly, if $W(\mathbf{d}|S_n,x)$ is 
bandlimited in $x$
and has the values $W_k$ at the equidistant sampled points $x_k$ then
$W(\mathbf{d}|S_n,x)$ is equal to $C(x)$. In the case where $\boldsymbol{\tau}$ is
multi-dimensional the interpolation can be performed separately on
each index: e.g., in the case of two dimensions [i.e.,
$\boldsymbol{\tau}$ equal to $(\tau_{1}, \tau_{2})$]
\begin{equation}
C(\boldsymbol{\tau}) = \sum_{j,k=-\infty}^\infty W_{jk}
\mathrm{sinc}{}\frac{\pi}{\Delta_1}\left(\tau_1-\tau_{1,j}\right)
\mathrm{sinc}\frac{\pi}{\Delta_2}\left(\tau_2-\tau_{2,k}\right),
\end{equation}
where
\begin{eqnarray}
\tau_{1,j} &=& \tau_{1,0} + j\Delta_{1},\\
\tau_{2,k} &=& \tau_{2,0} + k\Delta_{2}
\end{eqnarray}
and $\tau_{1,0}$, $\tau_{2,0}$ are constants. 

Cardinal interpolation using the Cardinal function $\mathrm{sinc}$
forms the basis of the interpolation formula used in
\cite{croce00a,croce00b}. If $W(\mathbf{d}|S_n,\boldsymbol{\tau})$ is bandlimited and
we choose our samples of $W$ appropriately then we can do no better
than using the Cardinal function to interpolate values of $W$ between
the samples. In our problem, however, $W(\mathbf{d}|S_n,\boldsymbol{\tau})$ is not
bandlimited and we do not have an infinite number of sample points
$W_k$; correspondingly, the Cardinal function $C(\boldsymbol{\tau})$
is at best an approximation to $W(\mathbf{d}|S_n,\boldsymbol{\tau})$. With that
understanding the Cardinal interpolation $C(\boldsymbol{\tau})$ is not
preferred and we are led to seek other approximations to
$W(\mathbf{d}|S_n,\boldsymbol{\tau})$ that have favorable properties\footnote{In
fact, as noted in [13,14], the $\Gamma$ is \emph{quasi-band-limited}:
i.e., there exists a $B_c$ such the error one makes by  
undersampling at frequency $B >B_c$ is proportional to $\exp[-(B-B_c)]$.
Nevertheless, interpolation with the Cardinal function is still an  
approximation and, as we are about to see, other interpolating  
functions can achieve equivalent accuracy at smaller computational costs.}.

One possibility, chosen from approximation (as opposed to
interpolation) theory, is the use of a Chebyshev polynomial expansion
to approximate $W(\mathbf{d}|S_n,\boldsymbol{\tau})$. Without loss of generality
consider a continuous function $f(x)$ on $[-1,1]$. The Weierstrass
Approximation Theorem states that for any $\epsilon>0$ we can find a
polynomial $P_n$ of order $n$ such that
\begin{equation}
\max_{x\in[-1,1]}\left|f(x)-P_n(x)\right|\leq\epsilon.
\end{equation}

The minimax polynomial approximation to $W(\mathbf{d}|S_n,x)$ is a natural candidate
for the interpolation $\widetilde{W}(x)$. Unfortunately, finding the
minimax polynomial is a very difficult process; nevertheless an
excellent \emph{approximation} to the minimax polynomial does exist.
Define the error $E(x|f,P_n)$ associated with the polynomial
approximation $P_n(x)$ by
\begin{equation}
E(x|f,P_n) \equiv f(x)-P_n(x).
\end{equation}
The Chebyshev  Equioscillation Theorem \cite{mason03a} 
states  $P^{*}_n$ is the minimax polynomial if and only if
there exist $n+2$ points $-1\leq x_0<x_1<\cdots<x_{n+1}\leq1$ for which 
\begin{equation}
E(x_k|f,P^*_n) = (-1)^kE,
\end{equation}
where
\begin{equation}
|E| \equiv \max_{x\in[-1,1]}\left|E(x|f,P_n)\right|.
\end{equation}
As a corollary, $E(x|f,P^{*}_n)$ vanishes for $x\in[-1,1]$ at $n+1$
points $x'_k$, with $x_k<x'_k<x_{k+1}$. This result, together with the
Mean Value Theorem, allows us to write the error term associated with
the minimax polynomial $P^{*}_n$ as
\begin{equation}
E(x|f,P^{*}_n) =
\frac{f^{(n+1)}(\xi)}{(n+1)!}\prod_{k=0}^n(x-x'_k), \label{def:E(x|f,P_n)} 
\end{equation}
where $\xi\in[-1,1]$. Correspondingly, 
\begin{equation}
|E| \leq \max_{x\in[-1,1]}\left|\prod_{k=0}^n(x-x'_k)\right|\,
\max_{\xi_\in[-1,1]}\frac{ \left|f^{(n+1)}(\xi)\right|}{(n+1)!}.
\end{equation}

Focus attention on the order $n+1$ polynomial
\begin{equation}
Q^{*}_{n+1}(x) = \prod_{k=0}^n(x-x'_k). 
\end{equation}
This polynomial has leading coefficient unity.  A unique property of
the Chebyshev polynomial $T_{n+1}$ is that, of all order $n+1$
polynomials $Q_{n+1}$ with leading coefficient unity,
\begin{equation}
\max_{x\in[-1,1]}\left|\frac{T_n(x)}{2^{n-1}}\right| \leq
\max_{x\in[-1,1]}|Q_{n+1}(x)|. 
\end{equation}
Additionally, $T_{n+1}(x)$ has exactly $(n+2)$ extrema on $[-1,1]$,
the value of $|T_{n+1}(x)|$ at these extrema is $1$, and the extrema
alternate in sign.  Correspondingly, if the error term
$E(x|f,P^{*}_n)$ associated with the minimax polynomial $P^{*}_n$ were
polynomial --- i.e., $f^{(n+1)}(\xi)$ were constant in equation
\ref{def:E(x|f,P_n)} so that $E(x|f,P^{*}_n)$ was equal to $Q^{*}_n$
--- then by the Equioscillation Theorem $Q^{*}_{n+1}$ would be equal
to $T_{n+1}$ and the $x'_k$ --- where the error vanishes --- would be
the $n+1$ roots of $T_{n+1}$. This suggests that we find the order $n$
polynomial $p_n^{*}$ such that
\begin{equation}
p^{*}_n(x'_k) = f(x'_k) \quad\forall \ k=0\ldots n
\end{equation}
where, again, the $x'_k$ are the roots of $T_{n+1}$.  The polynomial
$p^{*}_n$ is a \emph{near minimax} polynomial approximation to $f(x)$.
For this polynomial approximation Powell \cite{powell67a} showed that,
as long as $f(x)$ is continuous on $[-1,1]$,
\begin{equation}
1\leq
\frac{\epsilon_{\text{cheb}}}{\epsilon_0}
\leq \nu_n \equiv
1+\frac{1}{n+1}\sum_{k=0}^n\tan\left[\frac{(k+1/2)\pi}{2(n+1)}\right].
\end{equation}
where
\begin{eqnarray}
\epsilon_0 &=& \max_{x\in[-1,1]}\left|E(x|f,P^*_n)\right|,\\
\epsilon_{\text{cheb}} &=&  \max_{x\in[-1,1]}\left|E(x|f,p^*_n)\right|.
\end{eqnarray}
Powell also showed that $\nu_n$ grows slowly with $n$: in particular, 
\begin{equation}
\nu_n\sim\frac{2}{\pi}\log n.
\end{equation}
Somewhat tighter bounds on $\nu_n$ can be placed when $f$ is also
differentiable \cite{li04a}.

As defined above, the near minimax polynomial $p^{*}_n$ is the
interpolating polynomial that agrees with $f$ at the $n+1$ roots of
$T_{n+1}$.  Alternatively, using several properties of Chebyshev
polynomials, the Chebyshev interpolating polynomial can be expressed
as a linear combination of Chebyshev polynomials:
\begin{equation}
p^{*}_n(x) = \sum_{k = 0}^{n} a_k T_k (x) - \frac{1}{2} a_0,  
\end{equation}
where
\begin{equation}\label{eq:interpCoeff}
a_j = \frac{2}{n + 1} \sum_{k = 1}^{n + 1} f(x'_k) T_j (x'_k),
\end{equation}
where, again, the $x'_k$ are the $n+1$ roots of $T_{n+1}$.

\section{Comparison: Dense and Interpolated Search}\label{sec:Comparison}

In this section we describe the dense and interpolating search strategy and 
compare their efficiency when applied to the problem of identifying the 
gravitational wave signature of coalescing neutron star systems in the LIGO 
detectors. 

\subsection{Two Search Strategies}

The conventional search strategy used in the current analyses of LIGO, GEO 
and TAMA data (cf.\ \cite{sathyaprakash91a,dhurandhar94b,owen96a,LSC04d}) 
begins with the placement of templates at discrete points $\boldsymbol{\tau}_k$ 
on the parameter space $\boldsymbol{\tau}$. To choose the template locations we 
define the inner product of two signals $g(t)$ and $h(t)$, 
\begin{equation}
\left<g,h\right> = 4\int_0^\infty df\,
\Re\left[\dfrac{\widetilde{g}(f)\widetilde{h}^*(f)}{S_n(f)}\right],
\end{equation}
where $\widetilde{g}(f)$ is the Fourier transform of $g$ and $S_n$ is the 
detector noise power spectral density. Denoting by $h(t|\boldsymbol{\tau})$ 
the signal characterized by $\boldsymbol{\tau}$ the match
$\Gamma(\boldsymbol{\tau}_j,\boldsymbol{\tau}_k)$ is 
\begin{equation}
\Gamma(\boldsymbol{\tau}_j,\boldsymbol{\tau}_k)
= 
\frac{\left<h(t|\boldsymbol{\tau}_j),h(t,\boldsymbol{\tau}_k)\right>}{
\sqrt{\left<h(t|\boldsymbol{\tau}_j),h(t,\boldsymbol{\tau}_j)\right>
\left<h(t|\boldsymbol{\tau}_k),h(t,\boldsymbol{\tau}_k)\right>}}.
\end{equation}
By construction $|\Gamma|\leq1$. The templates locations are chosen so that 
consecutive templates in any of the 
directions $\tau_j$ have an overlap $\Gamma_0$, referred to as the ``minimum match'' 
and typically chosen to be 97\%. 

With the templates placed, the dense search strategy proceeds:
\begin{enumerate}
\item Evaluate the Wiener filter $W(\mathbf{d}|S_n,\boldsymbol{\tau}_k)$
at each of the template locations $\boldsymbol{\tau}_k$;
\item Determine the template $\boldsymbol{\tau}_j$ whose Wiener filter
output is greatest;
\item If the filter output at $\boldsymbol{\tau}_j$ exceeds the given threshold, 
report an event with the parameters $\boldsymbol{\tau}_j$. 
\end{enumerate}
We refer to this as the dense search strategy.

Following the discussion in section \ref{gen_method} we are in a position to describe an 
alternative strategy, which we refer to as the interpolated search strategy. First, 
fix the order of the interpolating polynomial. This determines the
template locations $\boldsymbol{\tau}_k$ 
on the parameter space $\boldsymbol{\tau}$. Then
\begin{enumerate}
\item Evaluate the Wiener filter $W(\mathbf{d}|S_n,\boldsymbol{\tau}_k)$
at each of the template locations $\boldsymbol{\tau}_k$;
\item Form the interpolating polynomial from the $W(\mathbf{d}|S_n,\boldsymbol{\tau}_k)$;
\item Determine the location $\boldsymbol{\tau}'$ where the interpolating polynomial 
is maximized;
\item Perform a final Wiener filter evaluation at $\boldsymbol{\tau}'$; 
\item If the final evaluation exceeds the given threshold, report an event with the parameters $\boldsymbol{\tau}'$. 
\end{enumerate}

\begin{figure}
\centering
\includegraphics[height=\textwidth, angle=-90]{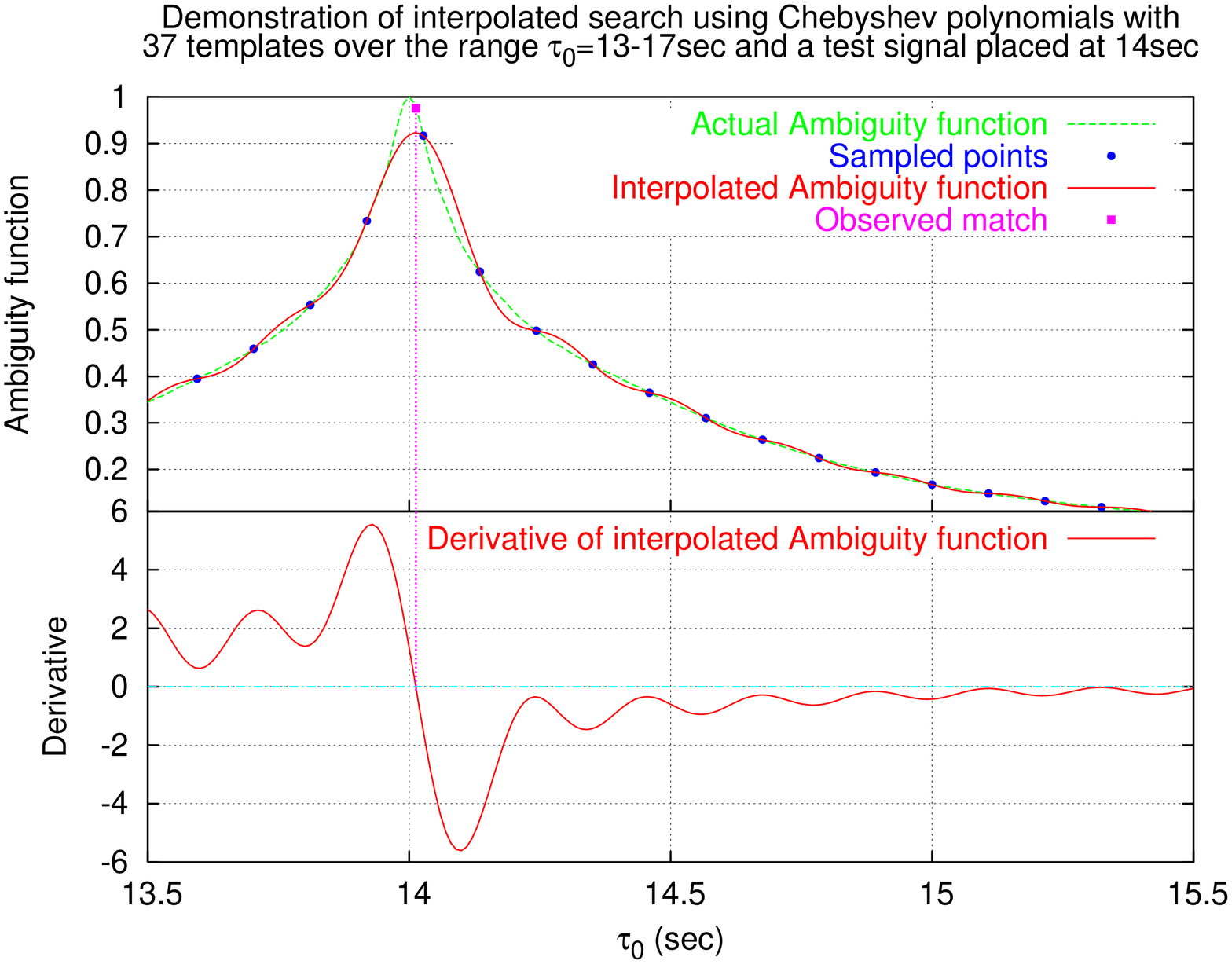}
\caption{This figure demonstrates the interpolated search. The ambiguity function is sampled and reconstructed over the chosen parameter space of $\tau_0=13-17$ sec (only a part of the parameter space is shown) with the help of the Chebyshev interpolating polynomial. The approximate location of the peak of the interpolating function is first located and the zero of the derivative is obtained by applying successive approximations around the peak. Note that by placing a template at the maximum of the interpolating polynomial, the match has improved over the one obtained by simply evaluating the maximum of the interpolating polynomial.}
\label{fig:demo}
\end{figure}

We illustrate the interpolated search strategy using Fig.~\ref{fig:demo}. In Fig.~\ref{fig:demo} we use $37$ interpolating search templates, that is, we sample the ambiguity function at $37$ points in the $\tau_0$ space (the marked points on the dotted curve). We construct the interpolating function (the solid curve) and find its maximum by setting its derivative to zero. In order to avoid local extrema, we first find the approximate location of the peak of the interpolating function and then find the zero of its derivative by successive approximation near the region of the peak. One can clearly see that the proper value of the ambiguity function at the maximum of the interpolating function is more than the maximum value of the interpolating function and this is what we gain by placing a template at the maximum of the interpolating function.

\subsection{A one-dimensional parameter space for comparative studies}

We are interested in understanding the performance of the 
interpolated search strategy relative to the dense search 
strategy, which is currently used in the analysis of data from the LIGO, 
GEO and TAMA detectors \cite{LSC04d}. The current analyses focus
on templates corresponding to binaries with circular orbits and no 
component spins. The corresponding two-dimensional parameter space is 
spanned by the masses of the individual components. The templates
vary most rapidly, however, along the axis spanned by 
the so-called \emph{chirp mass}
\begin{equation}
\mathcal{M} := \mu^{3/5}M^{2/5},
\end{equation}
where $M$ is the system's total mass and $\mu$ its reduced mass. 
The linear density of templates needed by the dense search in the 
direction $\partial_{\mathcal{M}}$ is approximately 100 times the linear
density needed in the orthogonal direction. For the comparison we 
perform here we focus attention on the number of template evaluations 
needed for binaries with equal mass components that vary only in 
$\mathcal{M}$. 
We expect
that the ratio of performance, measured as the number of templates
required by the two search strategies to achieve the same search 
results, will be the same in the complementary dimension and in 
the other dimensions that will be introduced in future searches 
that accommodate component spins and orbital eccentricity. 

\subsection{Templates}

The strain response of an interferometric gravitational wave detector to 
quadrupole formula approximation gravitational 
waves incident from an inspiraling binary neutron star system can be written
\begin{subequations}
\begin{equation}
h(t|t_a,\tau_0) = h_0 \left[\pi f(t-t_a-\tau_0)\mathcal{M}\right]^{2/3}\cos\Phi(t-t_a-\tau_0),
\end{equation}
where
\begin{eqnarray}
f(t|t_a,\tau_0) &:=& \dfrac{1}{\pi\mathcal{M}}
\left(\dfrac{5}{256}\dfrac{\mathcal{M}}{\tau_0+t_a-t}\right)^{3/8},\\
\Phi(t|t_a,\tau_0) &:=& \Phi_a + 2\pi\int_t^{t_a+\tau_0} dt\,f(t|t_a,\tau_0)
\end{eqnarray}
\end{subequations}
for $t<t_a+\tau_0$. Here $t_a$ is the moment when the instantaneous 
wave frequency is equal to $f_a$ and $\tau_0$ is the elapsed time
from that moment until (in this approximation) the system coalesces,
which is directly related to the system's chirp mass $\mathcal{M}$:
\begin{equation}
\tau_0 = \dfrac{5}{256\pi f_a}\dfrac{1}{\left(\pi\mathcal{M}f_a\right)^{5/3}}.
\end{equation}

The elapsed time to coalescence $\tau_0$ is a useful surrogate for the 
chirp mass $\mathcal{M}$: templates equispaced in $\tau_0$ have 
constant cross-correlation, independent of $\tau_0$. Choosing $f_a$
equal to 40~Hz, which is commonly taken as the lower-edge of the LIGO
detector bandwidth at design sensitivity \cite{LIGOE95001802}, $\tau_0$
ranges from approximately 43~s for a binary system consisting of 
two $1\,\text{M}_{\odot}$ compact objects to 0.15~s for
a binary consisting of two $30\,\text{M}_{\odot}$ black holes. 

It is convenient to work with the Fourier transform of the strain response 
of the detector. For neutron star binaries in the LIGO or Virgo band the
Fourier transform can be evaluated to an excellent approximation using
the stationary phase approximation \cite{finn93a}:
\begin{subequations}
\begin{equation}
\tilde{h}(f) = \mathcal{N}f^{-7/6}\exp\left\{i\left[-\Phi_a-\pi/4
+\Psi(f|t_a,\tau_0)\right]\right\},
\end{equation}
where
\begin{eqnarray}
\Phi_a &=& \Phi(t_a|t_a,\tau_0),\\
\Psi(f|t_a,\tau_0) &=& 2\pi f t_a + f_a\tau_0\dfrac{6\pi}{5}\left(\dfrac{f}{f_a}\right)^{-5/3}.
\end{eqnarray}
\end{subequations}
The factor $\mathcal{N}$ is a constant amplitude. 

It is important to distinguish between the nature of the parameters that
characterize the template. Changes in the parameter $\tau_0$ change 
the waveform \emph{shape}: we term such parameters \emph{dynamical}
parameters. On the other hand, parameters such as $t_a$ or $\Phi_a$
translate the waveform, but do not alter its shape: we term these
\emph{kinematical} parameters. In our problem only the subspace of dynamical 
parameters needs to be spanned by discrete templates: the values of the kinematic
parameters for the Wiener filter with the maximum output can be determined
by other means. Correspondingly, at the level of approximation associated with
the quadrupole formula the family of templates that must be evaluated is one
dimensional.

\subsection{Dense search template placement}

\begin{figure}
\centering
\includegraphics[height=0.7\textwidth,angle=-90]{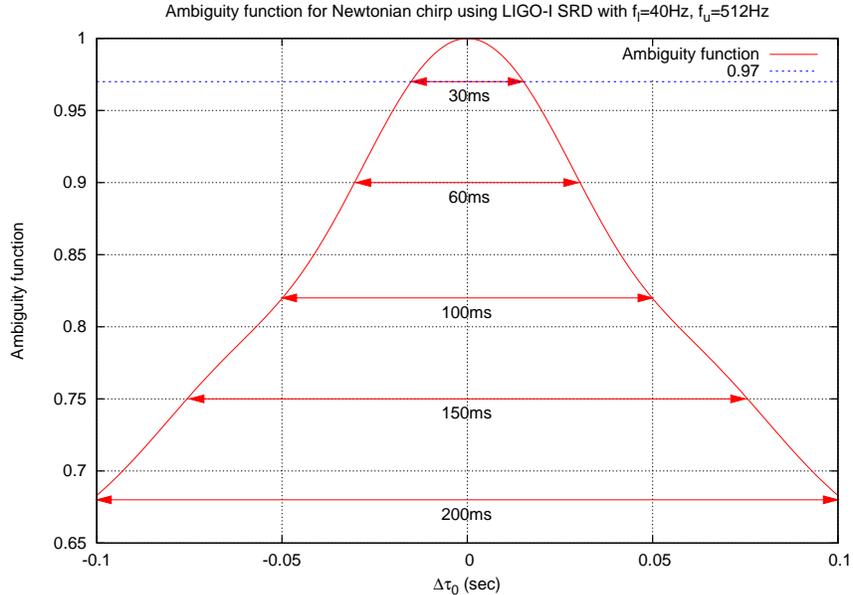}
\caption{Ambiguity function $\H$ plotted as a function of $\D \tau_0$. 
Horizontal lines are drawn for various matches; the dotted horizontal line 
is for a 97\% match, which corresponds to an inter-template separation 
of $\sim 30$ ms.}\label{fig:amb}
\end{figure}

There are many different ways of parameterizing the template 
space. Choosing $\tau_0$ as a dynamical variable has the advantage that 
$\Gamma(\tau_0,\tau'_0)$ depends only on the difference $\tau_0-\tau'_0$; 
consequently, in the dense search templates are spaced uniformly in 
$\tau_0$ \cite{sathyaprakash91a,dhurandhar94b,owen96a}.
To determine that spacing we evaluate 
\begin{equation}
\mathcal{H}(\Delta\tau_0) = \Gamma(\tau_0,\tau_0+\Delta\tau_0),
\end{equation}
where now $\Gamma$ has been maximized over the kinematical parameters $t_a$ and $\phi_a$. This maximization can be performed in a computationally efficient manner as shown in the literature \cite{sathyaprakash91a}.
We call $\mathcal{H}$ the dynamical ambiguity
function or simply as the ambiguity function. It quantifies the
fractional match between the template at $\tau_0$ and the signal at
$\tau_0 + \D \tau_0$. Figure \ref{fig:amb} shows $\mathcal{H}$ for 
power spectral density specified in the initial LIGO science requirements 
\cite{LIGOE95001802}. The requirement that $\mathcal{H}(\Delta\tau_0)$ is 
equal to a constant for any two consecutive templates determines the 
spacing $\Delta\tau_0$ between templates that differ only in $\tau_0$. 
For our example problem, which has just one dynamical parameter, 
the requirement that $\mathcal{H}(\Delta\tau_0)$ is 97\% (the conventional
choice) for neighboring 
templates leads to a template spacing $\Delta\tau_0$ equal to 30~ms. 

\subsection{Interpolated search template placement}

In the dense search templates are equispaced in $\tau_0$, with the spacing
between adjacent templates --- and thus the number of templates --- 
chosen such that the dynamical ambiguity function takes on a specified value. 
When presented with data an event is signaled when the amplitude at one of 
these templates exceeds a threshold. 

In the interpolated search, on the other hand, the domain $[\tau_0^{\min},\tau_0^{\max}]$
is mapped onto $[-1,1]$ and the 
placement and number of 
templates is chosen to simplify the construction of the Chebyshev interpolating
polynomial of the template output over this domain. When presented with data the 
maximum value of the Chebyshev interpolating polynomial is found and an event 
is signaled when the amplitude at that location exceeds a threshold. 

In the interpolated search our goal is to minimize the order of the 
interpolating polynomial (and, thus, the number of template evaluations) required 
for a given accuracy of interpolation. We have some control over this through the 
choice of mapping from $[\tau_0^{\min}, \tau_0^{\max}]$ to $[-1,1]$. The linear map
\begin{equation}
\tau' = 2\dfrac{\tau_0-\tau_0^{\min}}{\tau_0^{\max}-\tau_0^{\min}}-1,
\end{equation}
is the most obvious such mapping. While have not made an exhaustive search 
of all possible mappings; however, we have observed that better fits are possible 
with a lower-order polynomial when we use the mapping 
\begin{equation}\label{eq:delta}
\delta = - \cos \left[ \pi \, \dfrac{\tau_0- \tau_0^{\min}}{\tau_0^{\max}-\tau_0^{\min}} \right]. 
\end{equation}
Moreover, with this mapping, the roots of the Chebyshev polynomial are equi-spaced over the parameter range in $\tau_0$.
Once we have fixed the order $n$ of the interpolating polynomial templates 
are placed at values of $\delta$ that are roots of the $T_{n+1}(\delta)$ and 
the coefficients of the interpolating polynomial are found using 
equation~\ref{eq:interpCoeff}.

\begin{figure}
\centering
\includegraphics[width=\textwidth]{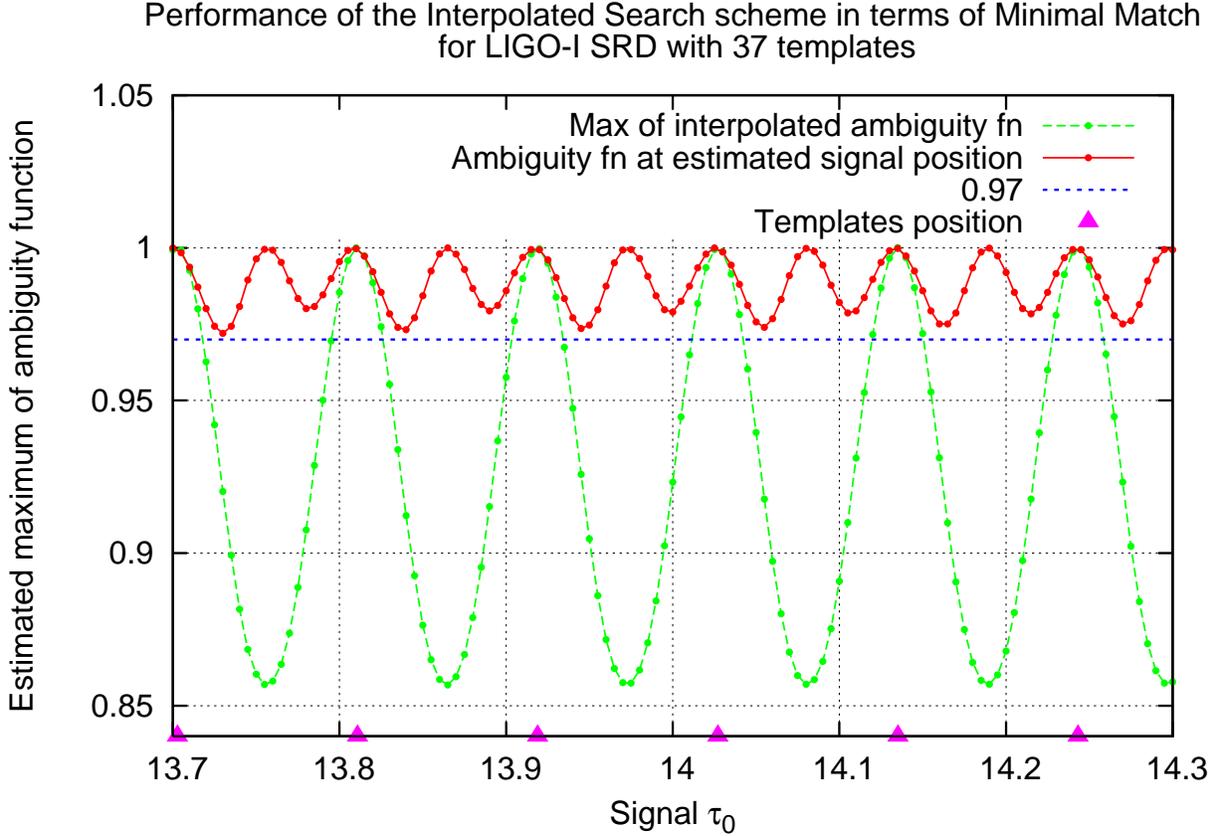}
\caption{Normalized test signals (without noise) were injected densely at regular intervals along the parameter space and the match obtained by the interpolated search strategy is plotted against $\tau_0$. This figure demonstrates that the match is a (nearly) periodic function of $\tau_0$ with the period equal to the template separation. Moreover, with just $37+1$ interpolated search templates the minimal match is $0.97$. For the same minimal match $133$ templates are needed for the usual dense search.}
\label{fig:MM}
\end{figure}

In Fig.~\ref{fig:MM} we have plotted the match by placing normalized test signals (without noise) at regular intervals of $\tau_0$. We see that the match is a (nearly) periodic function of $\tau_0$, with the period equal to the template separation. This suggests that the detection probability is also periodic and this fact has been used in carrying out the simulations - the signals are injected within one such ``period" in the parameter space. Moreover, one can see that with just $37+1$ interpolated search templates one gets a minimal match of $0.97$, whereas the dense search requires about $133$ templates to achieve the same level of minimal match. This amounts to a factor of $3.5$ over the dense search and this is so for just one dimension.

\subsection{Comparison}

We are interested in two, related, comparisons: first, the relative ``sensitivity'' of 
a search carried-out with a fixed number of template evaluations using the 
dense search strategy and the interpolated search strategy; second, the number 
of template evaluations required by the interpolated search in order to achieve 
the same ``sensitivity'' as the dense search. To give meaning to the ``sensitivity" 
of these two strategies we use the Receiver Operating Characteristic, or ROC. 

The ROC is a plot of true positives as a 
function of the fraction of false positives for a binary classifier system as 
its discrimination threshold is varied. Both the dense search and the interpolated 
search are binary classifiers: i.e., they classify an interval of data $\mathbf{d}$ 
as including a signal or not including a signal. A true positive is a
classification of $\mathbf{d}$ as including a signal when in fact it does; 
a false positive is a classification of $\mathbf{d}$ as including a signal
when it does not. In both of the search strategies described here the 
discrimination threshold is matched filter output that must be 
exceeded for a data interval to be classified as including a signal. The 
false positive fraction is also known as the type II, or false alarm, error fraction 
and is denoted $\alpha$. The fraction of true positives is also known as the 
detection efficiency $\epsilon$, which is one minus the type I, or false positive, 
error fraction (which is denoted $\beta$). At fixed $\alpha$ a more sensitive
search method has a greater $\epsilon$. The ROC associated with a search 
method no better than a toss of a (possibly loaded) coin is given by the diagonal 
$\alpha=\epsilon$. 

Using numerical simulations we have evaluated $\alpha$ and $\epsilon$ as a 
function of the detection threshold for both the interpolated search and the dense
search, for different numbers of templates (dense search) and different
interpolating polynomial order (interpolated search). 

To evaluate the false positive fraction $\alpha$ we generate a large number of data 
segments, each $2^{15}$ samples long, and each consisting of Gaussian noise 
whose power spectrum (assuming a 1024~Hz sample rate) is that specified as the 
initial LIGO science requirement \cite{LIGOE95001802}. (The Gaussian random numbers are 
themselves generated using the Mersenne Twister Pseudo Random Number 
Generator \cite{matsumoto98a} and then filtered in the Fourier domain by scaling the 
Fourier components by the square root of the PSD.) For the purpose of this comparison 
we look for signals in the interval $\tau_0\in[13\,\text{s},17\,\text{s}]$. Both the dense and 
interpolated search methods are applied to this data. The ratio of the number of events 
signaled to the number of data segments examined as a function of the threshold $\eta$ 
is $\alpha$ for that threshold. Approximately 50,000 realizations of detector
noise are used to evaluate $\alpha$, which gives reliable results for 
$\alpha$ greater than approximately $10^{-3}$. 

To compute $\epsilon$, the true positive fraction, we proceed in a similar fashion. 
Now, however, with each noise instantiation we add a signal, with $\tau_0$ drawn 
uniformly and randomly from the interval covered by the search: i.e., 
$\tau_0\in[13\,\text{s},17\,\text{s}]$. In almost all cases $50,000$ realizations of 
detector noise plus signal  are used to evaluate the efficiency, which gives reliable results for efficiencies greater than approximately $10^{-3}$. However, for the flat search with $40$ templates and the interpolated search with $30$ templates,we have used $400,000$ realizations. The larger number of realizations in these cases results in smoother curves.

\begin{figure}
\centering
\includegraphics[height=0.7\textwidth, angle=-90]{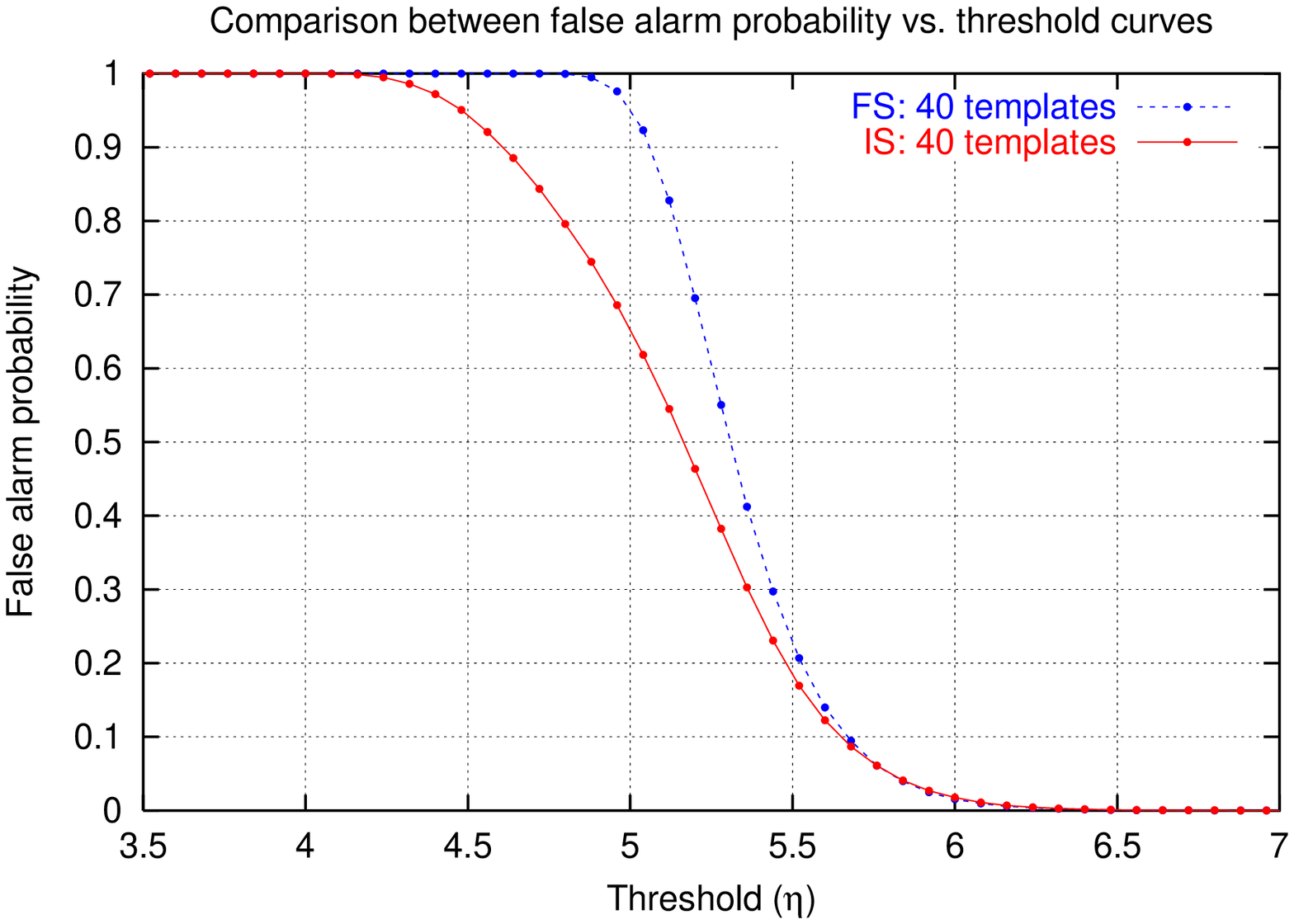}
\includegraphics[height=0.7\textwidth, angle=-90]{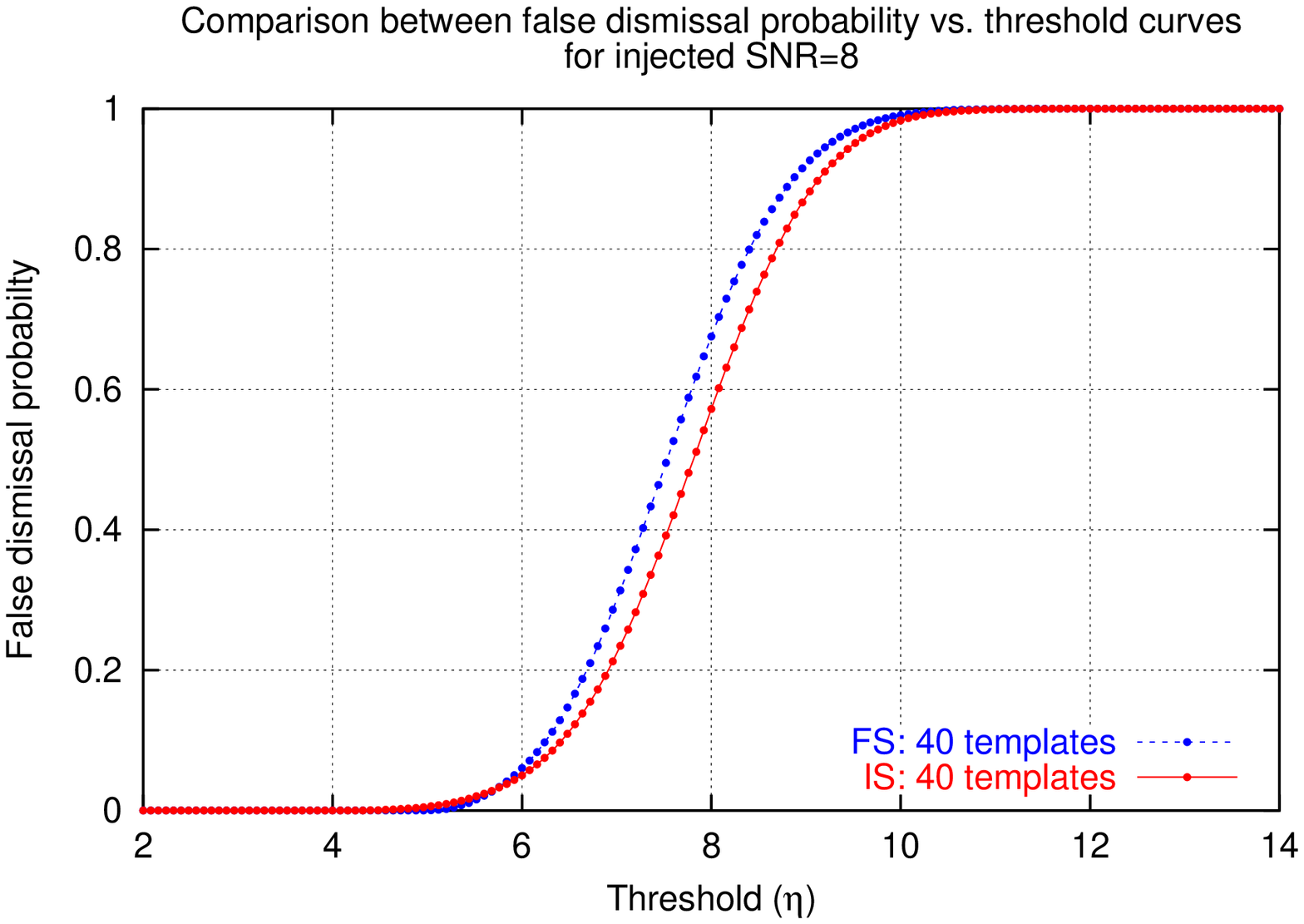}
\caption{The variation of the
false and true positive fractions, $\alpha$ and $\epsilon$ 
with threshold $\eta$ for the 
dense and interpolated search methods, each making use of 40
template evaluations. The top panel shows the false positive 
fraction. Note how the false positive falls much sooner for the 
interpolated search than for the dense search. The bottom panel
shows $\epsilon$ when a signal of amplitude
signal to noise 8 is present in the range $\tau_0\in[13\,\text{s},17\,\text{s}]$. 
Note how the $\epsilon$ is always greater for the 
interpolated search than for the dense search. For the same 
computational cost (determined by the number of template
evaluations) the interpolated search will always perform better 
than the dense search.} 
\label{fig:FA-Det}
\end{figure}

The top panel of Fig.~\ref{fig:FA-Det} shows the variation of  $\alpha$ for both methods 
using $40$ templates: i.e., a 100~ms template spacing for the dense search and
an order 39 interpolating polynomial in $\delta$ (cf.~equation \ref{eq:delta}). For
any threshold $\alpha$ is always greater for the dense search than for the interpolated 
search; similarly, as shown in the center panel of Fig.~\ref{fig:FA-Det}, for any given 
threshold the efficiency $\epsilon$ is always greater for the interpolated search than for the dense 
search. Finally, the bottom panel of Fig.~\ref{fig:FA-Det} shows the ROC for 
a 40 template dense search and an order 39 interpolated search, both of which involve 
40 template evaluations to decide if a signal has been detected. Comparing both ROCs
it is clear that the interpolated search is more sensitive at any given $\alpha$ then the 
dense search. This is always true: i.e., for a fixed number of template evaluations the 
interpolated search will always have a better efficiency at a given $\alpha$ than the 
dense search, though as the number of templates grows large the fractional difference
in sensitivity will decrease. 

\begin{figure}
\centering
\includegraphics[height=0.67\textwidth,angle=-90]{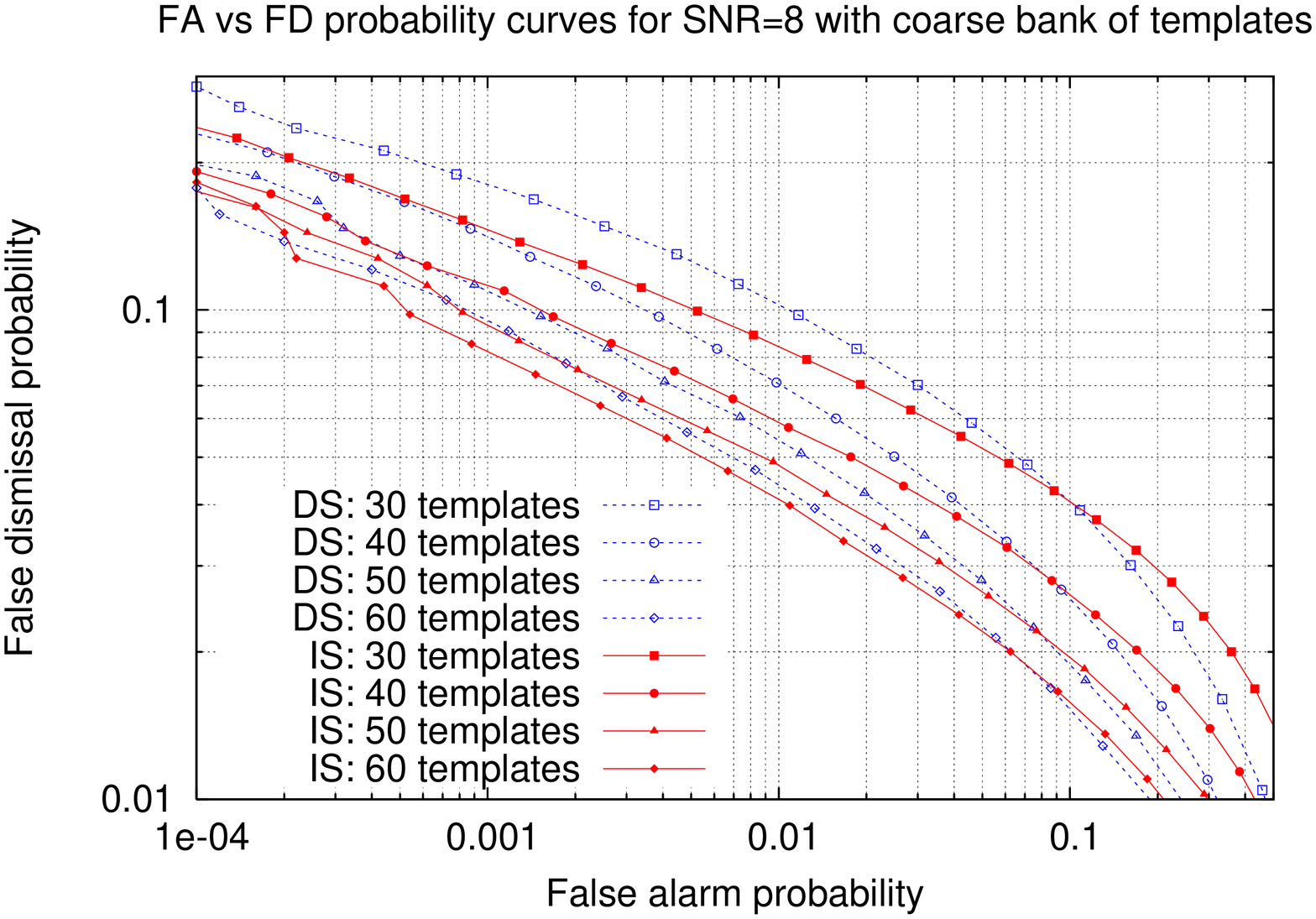}
\includegraphics[height=0.67\textwidth, angle=-90]{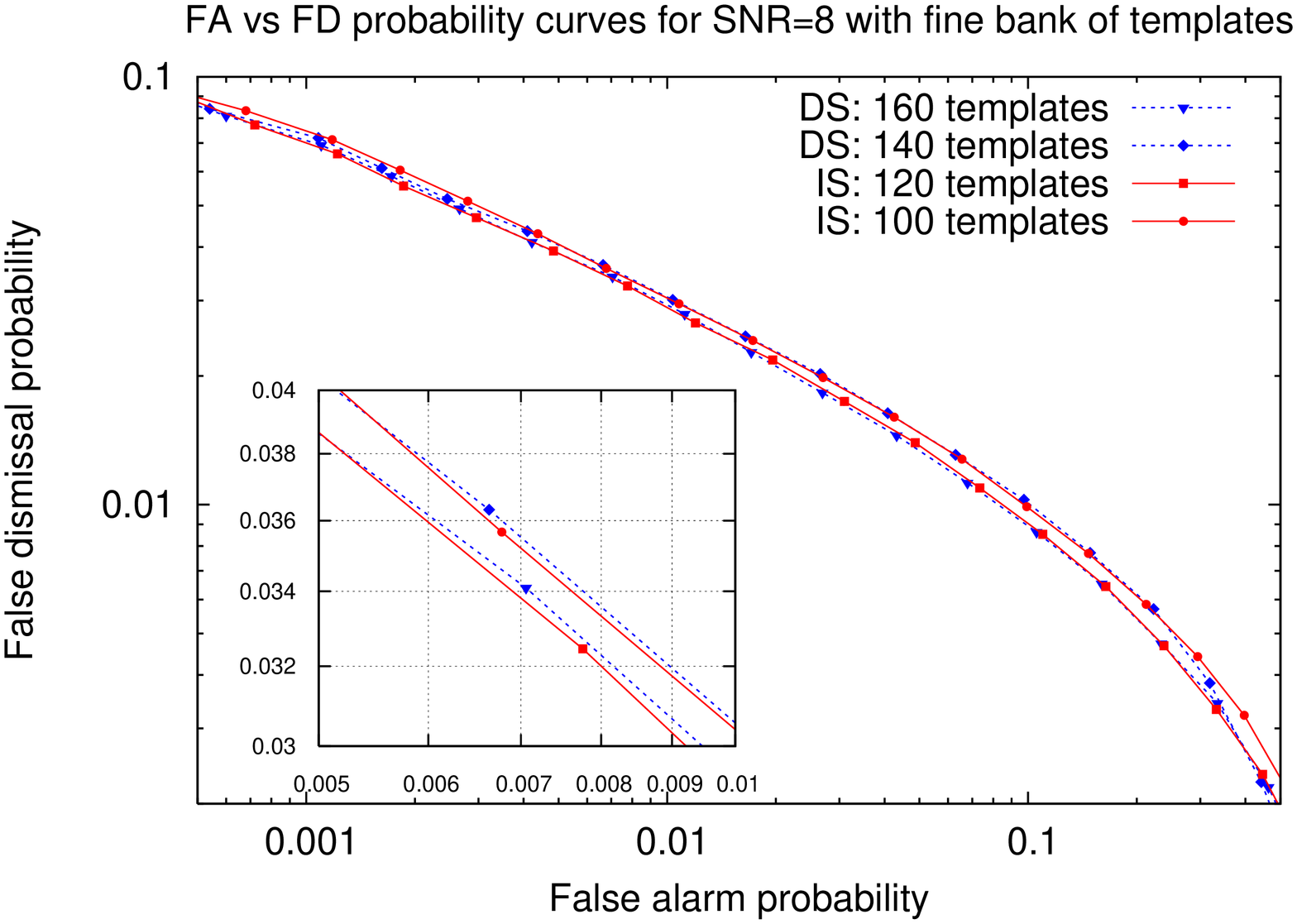}
\caption{ROC curves for dense searches (solid curves) and interpolating
searches (dashed curves). 
For a given number of templates, the solid curves are
  `lower' - less false dismissal probability for the same false alarm
  - than the dotted curves in the regime of low false alarm showing
  that the interpolated search performs better than the dense search
  for low false alarm. The bottom panel shows analogous plot for high
  minimal match (fine bank) $\sim .98$. Here the performance of the
  dense search with 160 and 140 templates is comparable to that of the
  interpolated search with 120 and 100 templates respectively.} 
\label{fig:sim}
\end{figure}

Figure~\ref{fig:sim} and table~I addresses the second of our two questions: the 
number of templates evaluations required of an interpolated search to 
have the same sensitivity as a dense search. Figure~\ref{fig:sim} shows the ROCs
for dense searches using 140 and 160 templates, together with the ROCs for 
interpolated searches using 120 and 100 templates. The interpolated search with
and order 120 interpolating polynomial is clearly as sensitive as a dense search with 
160 templates, and an interpolating search with an order 100 polynomial is as sensitive
as a dense search with 140 templates. Table~I shows similar pairings of the number of templates in a dense search and the number of templates in an interpolating search 
necessary to achieve the same sensitivity.

\begin{table}
\caption{Number of template evaluations required to obtain the same efficiency
at a false alarm fraction of $10^{-3}$ in a dense search and an interpolated search.
Note how the interpolated search is computational more efficient for the same
sensitivity.}\label{tbl:sim}
\begin{tabular}{|r|r|c|}\hline
\multicolumn{2}{|c|}{\# templates}&\multirow{2}{*}{$\epsilon$ at $\alpha=10^{-3}$}\\
\cline{1-2}
Dense&Interp.&\hfil{}\\
\hline
 40 &  31 & 0.859 \\ 
 50 &  41 & 0.890 \\ 
 60 &  49 & 0.905 \\ 
 80 &  64 & 0.919 \\ 
100 &  89 & 0.924 \\ 
140 & 105 & 0.927 \\ 
160 & 115 & 0.929 \\ 
\hline
\end{tabular}
\end{table}

\section{Conclusion}
\label{concl}

We have shown that the use of near-minimax interpolating polynomials 
to fit the output of matched filters to the filter parameter values can 
greatly improved the sensitivity of a matched-filter based search for
gravitational waves from compact binary inspiral. Using such a polynomial
to find the parameters of the signal template leading to the best match we 
can reduce the computational cost of a search over a two dimensional 
parameter space by a factor of two compared to the methods currently in use, 
without any loss of sensitivity or discriminating power. This factor of two becomes
a factor of ten when the search is over the seven dimensional parameter
space that includes not only the masses but also the spins of the binary 
components \cite{buonanno04a}. This savings in computational cost is
estimated under the assumption, which we believe well-founded, that  
we will obtain the same savings when the interpolation is extended to  
additional dimensions.

The use of near-minimax interpolation should be considered as part of
a larger strategy that employs a multi-grid approach to determine whether a signal 
has been observed and, if so, the parameters that characterize it. 
Since the major contribution to the computational cost of a multi-grid search is thought
to arise in the initial stage of the search the gain in computational efficiency --- 
and, correspondingly, 
the size of the parameter space that can studied with fixed computational resources
--- could be substantial. 

\section*{Acknowledgments}
\label{ack}

We would like to thank Innocenzo Pinto for his extremely useful comments.
We are grateful to Albert Lazzarini and Anand S. Sengupta for
valuable discussions. We are glad to acknowledge the use of Pleiades
cluster facility at Penn State, which is supported by Penn State University,
the Center for Gravitational Wave Physics, the International Virtual Data
Grid Laboratory, and the National Science Foundation (PHY-00-99559). 
We are also glad to thank IUCAA for the use of its high performance 
computing facility. SM acknowledges the support of CSIR.
SVD acknowledges the support of the DST. 
LSF acknowledges the support of the Center for Gravitational Wave Physics 
and the NSF under awards PHY-00-99559 and INT-01-38459. The 
International Virtual Data Grid Laboratory is supported by the NSF under 
cooperative agreement number PHY-01-22557; the Center for Gravitational 
Wave Physics is supported by the NSF under cooperative agreement 
PHY-01-14375.

\bibliographystyle{apsrev}

\end{document}